\documentstyle[aps]{revtex}

\begin{document}
\twocolumn [\hsize\textwidth\columnwidth\hsize
           \csname @twocolumnfalse\endcsname
\draft

\begin{center}
\large Isospin Influences on Particle Emission and Critical Phenomenon in Nuclear Dissociation
\end{center}

\begin{center}
\ Y.G.\ Ma$^{a,b,c}$ 
\footnotemark \footnotetext{\footnotesize Corresponding Author;  Email Address: mayugang@public1.sta.net.cn},
\ Q.M.\ Su$^b$, \ W.Q.\  Shen$^{b}$, \ D.D.\ Han$^d$, 
\\ \ J.S.\ Wang$^b$, \ X.Z.\ Cai$^b$, \ D.Q.\ Fang$^b$, 
\ H.Y.\ Zhang$^b$
\vspace{.5cm}

$^{a}$ CCAST (World Laboratory), P. O. Box 8730, Beijing 100080, China


$^{b}$ Shanghai Institute of Nuclear Research, Chinese Academy of Sciences,
P.O. Box 800-204, Shanghai 201800, China \footnotemark \footnotetext{Mailing address.}

$^{c}$ T.D.Lee Physics Laboratory, Fudan University, Shanghai 200433, China

$^{d}$ Electric Enginnering Dept., East China Normal University, Shanghai 200061, China
\end{center}
\maketitle
\begin{abstract}
Features of particle emission and critical point behavior are investigated as  functions of the isospin of disassembling sources and temperature at a moderate freeze-out density for medium-size Xe isotopes in the framework of isospin dependent lattice gas model. Multiplicities of  emitted light particles, isotopic and isobaric ratios of light particles show the strong dependence on the isospin of the dissociation source, but  double ratios of light isotope pairs and  the critical temperature determined by the extreme values of some critical observables are insensitive to the isospin of the systems. Values of the 
power law parameter of cluster mass distribution, mean multiplicity of intermediate mass fragments ($IMF$), information entropy ($H$) and Campi's second moment ($S_2$)  also show a minor dependence on the isospin of Xe isotopes at the critical point. In addition, the slopes of the average multiplicites of the neutrons ($N_n$), protons ($N_p$), charged particles ($N_{CP}$), 
 and IMFs ($N_{imf}$), slopes of the largest fragment mass number ($A_{max}$), and the excitation energy per nucleon of the disassembling source ($E^*/A$) to temperature are investigated as well as variances of the distributions of $N_n$, $N_p$, $N_{CP}$, $N_{IMF}$, $A_{max}$ and $E^*/A$.  
It is found that they  can be taken as additional judgements to the critical phenomena. 
\end{abstract}

\vskip2pc]

\narrowtext

\section{Introduction}

Isospin influence on the formation and decay of hot nuclei is an important subject in heavy ion collision physics nowadays.  Interests in this direction have been largely pushed away with the development of the radioactive beam technique. Many new impressive experiments aiming to explore such isospin effects can be performed by using the radioactive beams and/or targets with large neutron or proton excess. It offers the possibility to study the properties of nuclear matter in the range from symmetrical nuclear matter to pure neutron matter. Recently some theoretical investigations to the nuclear equation of state, chemical and mechanical instabilites as well as liquid-gas phase transition for isospin asymmetrical nuclear matter  were performed already \cite{Bali98}. In addition, the study on isospin dependent nucleon - nucleon cross section \cite{Teha94,Bert88,Ligq1,Cai98} is also an important subject due to its significant effect on the dynamical process of heavy ion reactions induced by radioactive beams. Experimentally some new isospin dependent phenomena  have been also discovered. For examples, the isospin dependences of preequilibrium nucleon emission \cite{Demp96,Lozh92,Kund96,Ba97},  nuclear stopping \cite{Yenn94}, nuclear collective flow \cite{Rpak97,Liba97},  total reaction cross section, radii of neutron-rich nuclei \cite{Shen89,Ygma93,Suzu95} and  subthreshold pion production \cite{Naga81} have been studied by several groups. However, more experimental and theoretical studies are still needed to better understand the isospin physics. 

On the other hand, phase transition and critical point behavior in a finite nuclear system is a debating topics. Two kind of important phase transitions of nuclear matter, namely the liquid gas phase transition and the quark-gulon plasma  phase transition which have been predicted to occur in HIC at intermediate  and ultra-relativistic energy domain, respectively, are attracting more and more nuclear scientists to pay attentions. Due to the manifest significance of both phase transitions in clarifing the properties of nuclear matter in extreme conditions, theoretical physicists are trying to present various possible judgements to characterize the phase transition with divers models as well as experimentalists are searching for such any evidences with the help of advanced and complicated set-up.  

In this article, we will investigate some isospin effects on cluster emission and critical point behavior stemming from  nuclear liquid gas phase transition in the framework of isospin dependent lattice gas model. The paper is organized as following: In Sec. II, a brief description of the isospin dependent lattice gas model is given. Results and discussions for the disassembly of $^{122,129,137,146}Xe$ isotopes in a fixed freeze-out density of $\sim$ 0.39$\rho_0$ are presented in Sec.III. Firstly, the multiplicities of particle emission are used to discuss the influence of the isospin. In addition,  the slopes of some average quantities to temperature and the variances of the distributions of some physical observables are explored to be the judgements to locate critical point. Secondly, the isospin dependence of the ratios between two light particles (LP) is investigated. Thirdly, the double ratios between two pair light isotopes are studied. Fourthly, the critical observables with different isospin are presented. Finally the summary  is given in Sec. IV.

\section{Description of Model}

The lattice gas model of  Lee and Yang \cite{Yang52}, where the grancanonical partition function of a gas with one type of atoms is mapped into the canonical ensemble of an Ising model for 
spin 1/2 particles, has successfully described the liquid-gas phase transition for atomic system. The same model has already been applied to nuclear physics for isospin symmetrical systems in the grancanonical ensemble \cite{Biro86} with an approximate sampling of the canonical ensemble \cite{Mull97,Jpan95,Jpan96,Camp97,Gulm98}, and also for isospin asymmetrical nuclear matter in the mean field approximation \cite{Sray97}. In this article, we will adopt the similar lattice gas model developed by J. Pan and S. Das Gupta \cite{Jpan95,Jpan96}. Some details and features of the model can be found in their papers. In comparison with the earlier version of their model \cite{Jpan95,Jpan96}, the different points in this work are that the exact Metropolis sampling for placing nucleons on the cubic lattice are adopted and while the isospin dependent interaction potential  is taken (which has been also incarnated in their recent works \cite{Gupta97,Panj99}). To better understand the context of this work we will make a brief description for the model and Monte-Carlo Metrolpois sampling technique.
 
In the lattice gas  model, $A$ nucleons with an occupation number $s$ which is defined $s$ = 1 (-1) for a proton (neutron) or $s$ = 0 for a vacancy, are placed on the $L$ sites of lattice. Nucleons in the nearest neighboring sites have interaction with an energy $\epsilon_{s_i s_j}$. The hamiltonian is written by 

\begin{equation}
E = \sum_{i=1}^{A} \frac{P_i^2}{2m} - \sum_{i < j} \epsilon_{s_i s_j}s_i s_j
\end{equation}

The interaction constant $\epsilon_{s_i s_j}$ is fixed to reproduce the binding energy of the nuclei. In order to incorporate the isospin effect in the lattice gas model, the short-range interaction constant $\epsilon_{s_i s_j}$ is made the difference between the nearest neighboring like-nucleons and unlike-nucleons:    $\epsilon_{nn}$ = $\epsilon_{pp}$ =
 $\epsilon_{-1-1}$ = $\epsilon_{11}$ = 0. MeV, $\epsilon_{pn}$ = $\epsilon_{np}$ = 
 $\epsilon_{1-1}$ = $\epsilon_{-11}$ = - 5.33 MeV, which indicates the repulsion 
between the nearest neighboring like-nucleons and attraction between the
 nearest neighboring unlike-nucleons. This kind of isospin dependent interaction incorporates, in a certain extent, Pauli exclusion principal and avoids effectively to produce some unreasonable clusters, like di-proton and di-neutron etc. Three-dimension cubic lattice with $L$ sites is used which results that an assumed freeze-out density of disassembling system is $\rho_f$ = $\frac{A}{L} \rho_0$, in which $\rho_0$ is the normal nuclear density. The disassembly of the system is to be calculated at $\rho_f$, beyond which nucleons are too far apart to interact.  $N + Z $ nucleons are put into $L$ sites by Monte Carlo sampling using the exact Metropolis algorithm \cite{Mull97,Metr53}. 

As pointed out in Ref. \cite{Shida,Carmona}, one has to be careful to treat the process of Metropolis sampling to fulfill the detailed balance and therefore warrant the correct equilibrium distribution in final state.  Speaking in detail, in this work, first we establish an initial configuration with $N + Z$ nucleons. Second, for each event, we will test sufficient number of "spin"-exchange steps, eg. 20000  steps in this work to let finally  the system generates states with a probability proportional to the Boltzmann probability distribution with Metropolis algorithm. In each "spin"-exchange step, we make a random trial change on the basis of the previous configuration. For example choose a nucleon at random and attempt to exchange it with one of its neighboring nucleons or the vacancies at random regardless of the sign of its "spin" (Kawasaki-like spin-exchange dynamics \cite{Kawasaki}), then compute the change $\Delta E$ in the energy of the system due to the trial change. If $\Delta E$ is less than or equal to zero, accept the new configuration and repeat the next "spin"-exchange step. If $\Delta E$ is positive, compute the "transition probability" $W$ = $e^{-\Delta E/T}$ and  compare it with the a random number $r$ in the interval [0,1]. If $r$ $\leq$ $W$, accept the new configuration; otherwise retain the previous configuration, and then start the next "spin"-exchange step. 20000 "spin"-exchange steps are performed to assure to get the equilibrium state. Third, once the nucleons have been placed stably on the cubic lattice after 20000 "spin"-exchange steps for each event, their momenta are generated by  Monte Carlo sampling of Maxwell Boltzmann distribution. Thus various observables can be calculated in a straightforward fashion for each event. One point should be emphasized here is that the above Monte-Carlo Metropolis "spin"-exchange approach between the nearest neighbors, independent of their "spin", is evidenced to be satisfied to the detailed balance as noted in Ref. \cite{Shida,Carmona}. In other word this sampling method will guarantee that the generated microscopic states  form an equilibrium canonical ensemble.


One of the basic measurable quantities is the size distribution of clusters. The definition of clusters has been extensively discussed in the previous references \cite{Jpan95,Jpan96,Camp97}. The first method is so-called Ising cluster which combines all the connected sites as cluster. However, it is not proper approach to define clusters in the lattice gas model since it does not fulfill the requirement that the correlation length should diverge at the critical point \cite{Camp97}. In condensed matter physics, Coniglio and Klein \cite{Coni80} proposed to combine the above site percolation with an addition bond percolation algorithm using a temperature dependent bonding  probability $p(T) = 1 - e^{-\epsilon_{s_i s_j}s_i s_j/2T}$. In the lattice gas model a similar way to  Coniglio-Klein's prescription is extensively adopted to define the cluster \cite{Jpan95,Jpan96,Camp97,Gulm98}: $ie.$  two neighboring nucleons 
are viewed to be in the same fragment if their relative kinetic energy 
is insufficient to overcome the attractive potential:  
$P_r^2/2\mu - \epsilon_{s_i s_j}s_i s_j < 0 $. It results in a similar temperature dependent bonding probability to Coniglio-Klein's prescription.

\section{Results and Discussions}

Four isotopes of Xe are chosed as  examples to illustrate the isospin effect with the help of  the isospin dependent lattice gas model. Their isospin parameter ($I$ = $\frac{N-Z}{A}$) is
about  0.11, 0.16, 0.21 and 0.26 for $^{122}$Xe, $^{129}$Xe, $^{137}$Xe and $^{146}$Xe, respectively. The freeze-out density $\rho_f$ has been chosen to be close to 0.39 $\rho_0$,
as extracted from the analysis of Ar + Sc \cite{Jpan95} and $^{35}$Cl + Au and $^{70}$Ge + Ti \cite{Beau96} with the same model. There is also good support from experiment that the value
 of $\rho_f$ is significantly below 0.5$\rho_0$ \cite{Agos96}.  343  sites with $7\times 7 \times 7$ cube results in 0.36, 0.38, 0.40, and 0.43 $\rho_0$ of the freeze-out density $\rho_f$ for $^{122,129,137,146}$Xe, respectively. Calculations were performed  from 3 to 7 MeV with a step 0.5 MeV and 1000 events were accumulated at each temperature for each isotope.

\subsection{Cluster Emission and Its Application to Locate the Critical Temperature}

\subsubsection{Multiplicity of Clusters, the Largest Fragment Mass and Excitation Energy}
 
Before presenting the calculated results, we firstly introduce a definition of excitation energy in this lattice gas model. The excitation energy per nucleon  can be written as \cite{Jpan96}
\begin{equation}
E^*/A = E_T - E_{g.s.} = ( \frac{3}{2}T + \epsilon_{np} \frac{N_{np}^T}{A})
-\epsilon_{np} \frac{N_{np}^{g.s.}}{A} 
\end{equation}
in which  $N_{np}^{g.s.}$ and $N_{np}^T$  is the number of the bonds of unlike-nucleons in the ground state (zero temperature) and at $T$, respectively. 
Experimentally the excitation energy of a nuclear system is generally given with respect to a cold ($T$=0) nucleus at normal nuclear density. In this classical model, we adopt the similar definition for the ground state, $ie.$ it corresponds to a cold nucleus at zero temperature and normal nuclear density where there is no kinetic energy and so that the ground state energy per nucleon is $-\epsilon_{n,p} \frac{N_{n,p}^{g.s.}}{A}$. Practically, $N_{n,p}^{g.s.}$ is determined by the geometry and is taken as the maximum bond number of unlike nucleons for disassembling source as it approaches to zero temperature and normal nuclear density as possible. 

Figure 1 shows the average multiplicites of the emitted neutrons, protons, charged particles and intermediate mass fragments, the average  values of the largest fragment masses and the average excitation energies per nucleon of disassembling source as functions of temperature and isospin. In point  of temperature dependence, $N_n$, $N_p$, $N_{cp}$ and $E^*/A$ increase while $A_{max}$ decreases with temperature  as expected. However, $N_{imf}$ shows a rise and fall with temperature as seen in previous studies \cite{Ogil91,Tsang,Mayg95,Macpl1}, which has been explained by the onset of the multifragmentation. In point of isospin dependence, $N_n$ shows a positive correlation with $I$ which is obviously reasonable due to the more neutrons for isotopes with larger $I$. Similarly,  $N_{imf}$ and $A_{max}$ show minor positive correlation with $I$. Reversely, $N_p$, $N_{cp}$ and $E^*/A$ seem to have  anti-correlation with $I$ at the same temperature even though the same  number of protons begin with in  these  dissociation isotopic sources. Experimentally the similar pictures for free neutrons and protons were found in $^{32}$S + $^{144,154}$Sm reactions \cite{Hilscher}. How to understand these phenomena? 

Two interpretations seem to be possible. On one hand, symmetrical potential term in Eq.(1) will play an important role in controlling the emission of nucleons and clusters. With the increasing of the isospin of the sources, protons will feel stronger attractive potential due to the neighboring neutrons sharing alike increases which results in more bound protons for disassembling sources with higher isospin. Viceversa, neutrons will, on one side, feel the stronger repulsive interaction due to more neutron-neutron pairs and, on other side, feel  smaller attractive potential because of  the decreasing of the average assortative number of nearest neighboring protons for a certain neutron with the increasing of the isospin of sources, both reasons will lead to produce more unbound neutrons for disassembling sources with higher isospin. This explication is similar to the explanation to nucleon emission in isospin dependent transport model \cite{Bali98}. However, another interpretation based on the excitation energy seems to be possible, too. In a previous work  we explored that the excitation energy can be viewed as a scaling quantity to control nuclear disassembly \cite{Maepj}.  If we made the mapping from T to $E^*/A$ (see Fig.1f) and re-plot the $N_n$, $N_p$, $N_{cp}$, $N_{imf}$ and $A_{max}$  as a function of excitation energy per nucleon instead of temperature in  Figure 2. Clearly, the rule for average multiplicity of neutrons is not changed,  $ie.$ the higher the excitation energy and/or isospin,  the larger the $N_n$. But for protons, charged particles, intermediate mass fragments and the largest fragment masses, their average values for  four isotopes nearly merge approximately into single curves. In this case, the role of symmtrical potenial is transferred into the  excitation energy of different sources and the isospin dependence of $N_p$, $N_{cp}$, $N_{imf}$ and $A_{max}$ at the same temperature showed in Fig.1b-e is nothing but the excitation energy dependence. 

\subsubsection{Slopes of Average Values to Temperature}
In Fig.1 the average values of most physical quantities increase or decrease monotonically  with temperature except for the IMF. Is it possible to obtain much information from this figure? Figure 3 gives slopes of $N_n$, $N_p$, $N_{cp}$, $N_{imf}$, $A_{max}$ and $E^*/A$ to temperature. Obviously, each slope has a peak or valley around $T$ $\sim$ 5MeV. In such
turning temperature, some features appear: (1) the emission of light particles and  complex fragments increases rapidly within a narrow temperature range,  reflecting that  the phase space is opening in the largest extent in that time;  (2) the decrease of the largest fragment size reaches to  the valley value for such a finite system. Physically the largest fragment is simply related to the order parameter $\rho_l$ - $\rho_g$ (the difference of density in the
$"liquid"$ and $"gas"$ phases). In the infinite matter, the infinite cluster exists
only on the $"liquid"$ side of critical point. In a finite matter, the largest
cluster is present on both sides of critical point. In this calculation,
a valley for the slope of $A_{max}$ to temperature may correspond to a sudden
disappearance of infinite cluster ($"bulk\ liquid"$) near critical temperature;
(3) the specific heat $C_v/A$ ($ie.$ $\partial (E^*/A)/\partial T$) has a peak
value for such a finite system. All these features are consistent to the concept
of phase transition and critical phenomenon according to the thermodynamical
theory. Hence, these slopes can be viewed as a characteristic judgement to  critical
phenomenon as other critical observables can do (see Sec. D). Moreover, 
the fact that the turning temperatures deduced from these slopes for the four isotopic disassembling sources nearly locate at the same temperature of $\sim$ 5 MeV, 
independent of the  isospin, illustrates that the critical temperature is insensitive to the isospin of the disassembling sources. This conclusion is consistent to the results deduced from other critical observables as shown below (see Sec. D). 

\subsubsection{The Largest Fluctuations}
Furthermore, the largest fluctuations are also found around critical temperature in the same calculation. Fig.4 illustrates that RMS width ($\sigma$) for the multiplicity distributions of neutrons, protons, charged particles and intermediate mass fragments, for the distributions of the largest fragment masses and excitation energies per nucleon. The variances of multiplicity distributions of neutrons and protons have maximum values or tends to saturation around 5 $\sim$ 5.5 MeV, and the variances for the distributions of $CP$, $IMF$, $A_{max}$ and $E^*/A$ show peaks at the same critical temperature. The peaks in the  critical point correspond to the cusp of the variances, $ie.$ their first order derivations are discontinual across the critical temperature which is also an indication of phase transition \cite{Chas96}. Noting that the fluctuation of $A_{max}$ is related to the compressibility of the system. These features are consistent with the critical point behavior where there should be the largest fluctuation in terms of statistical theory. Similarly to the results of slopes, the deduced critical temperature from the largest fluctuation is almost the same regardless of the isospin of the dissociation sources.

\subsection{Ratios between Two Light Particles}

Besides the multiplicity of emitted particles, the ratio between two light particles is probably  suitable to investigate the isospin effect. Fig.5 shows the temperature dependences of the isotopic ratios: $R(p/d)$ between the yield of protons to that of deuterons and $R(d/t)$ between the yield of  deuteron to that of triton, the isobaric ratios: $R(t/^3He)$
between the yield of triton and that of helium-3 and $R(^6He/^6Li)$ between the multiplicity of helium-6 to that of lithium-6, and the ratios $R(^A_Z X /^{A+1}_{Z+1} Y)$
between the light particles having one proton and mass number difference: $R(d/^3He)$ between the multiplicity of deuterons and helium-3 and $R(t/^4He)$ between the multiplicity of triton to that of helium-4 for the disassembly of Xe. For the isotopic ratios, $R(p/d)$ and $R(d/t)$, 
the particle in  denominator has one more neutron than that in numerator,
$ie.$ the ratios could reflect the extent of neutron-poor of the products.
Clearly, these ratios decrease with the increasing of isospin as shown in Fig.5a and 5b. It is consistent with the N/Z systematics of the disassembling source. R(p/d) and R(d/t) exists a wide valley around 5 MeV and then  increases with temperature above 5 MeV (near the critical temperature), a similar result was observed experimentally in energetic Au + Au collisions \cite{Ahle98}. These results are qualitatively consistent with a rise and fall in the relative production of IMF as shown above (see Fig.1d) \cite{Ogil91,Tsang,Mayg95,Macpl1}. In a simple coalescence picture this is consistent with a lower baryon density at freeze-out for the most violent collision and  disassembly \cite{Ahle98}. For the isobaric ratios, $R(t/^3He)$ and $R(^6He/^6Li)$, the particle in denominator has one more proton than that in  numerator, $ie.$ the ratio will indicate the extent of neutron-rich of the products. Fig.5c and 5d illustrates that the larger the isospin, the higher the ratios. It has the same trend to N/Z of emitter, too. Their values of ratios become saturation at higher temperature. For the ratios of $R(d/^3He)$ and $R(t/^4He)$, the particle in denominator has the same neutron as that in numerator but has one more proton, $ie.$, this ratio can show the extent of proton-deficient or neutron-rich of products. Fig.5e and 5f show that these ratios can also imply the extent of isospin of the emitted source. The dependence of temperature is similar to $R(t/^3He)$ and $R(^6He/^6Li)$. Besides the above ratios, another understandable quantity is the ratio of yield 
of emitted neutron to protons, $R(n/p)$ (not plotted here because both multiplicities of neutrons and protons have been shown in Fig.1), which increases with the isospin of the source and tends to saturation at higher temperature.

In order to display the isospin systematics of the above ratios clearly, we plot the isotopic ratios of $R(p/d)$, $R(d/t)$, $R(^3He/\alpha)$ and $R(^6Li/^7Li)$ as a function of the isospin $I$ in a fixed temperature, eg. around critical temperature $\sim$ 5 MeV in Fig. 6a. Obviously all the ratios decrease with $(N-Z)/A$ which is consistent with the isospin systematics of disassembling source. 

\subsection{Double Ratios between Two Pair Light Isotopes}

After investigating the multiplicities of LP and the ratios between one pair LP, it is interesting to study the double ratios between two pair LP. In the assumption  of thermal and chemical equilibrium, S. Albergo $et\ al.$  \cite{Albe85} derived a formula of temperature by the double ratios of isotope. The double ratios of isotope can cancel out chemical potential effects and offer a particular promising technique of temperature determination
 \cite{Poch95,Haug96,Mayg97,Serf98,Mila98,Tsang97,Xi98,Xi97}.
 But due to side-feeding of light particles of excited primary fragments and other complicate effects in experimental measurements, the determination  of temperature from double ratio must be taken carefully \cite{Tsang97,Siwek98,Gulm97,Bondorf98}.
However,  here we restrict to discuss the double ratios between  different light isotopes rather than the isotopic temperature.

Fig.6b shows  the double ratios between two pair isotopic ratios, namely
$R(pd-He)$ = $\frac{N_{p}/N_{d}}{N_{^3He}/N_{^4He}}$,
$R(pd-Li)$ = $\frac{N_{p}/N_{d}}{N_{^6Li}/N_{^7Li}}$,
$R(dt-He)$ = $\frac{N_{d}/N_{t}}{N_{^3He}/N_{^4He}}$,
$R(dt-Li)$ = $\frac{N_{d}/N_{t}}{N_{^6Li}/N_{^7Li}}$,
and 
$R(He-Li)$ = $\frac{N_{^3He}/N_{^4He}}{N_{^6Li}/N_{^7Li}}$. 
Different from the isotopic ratios of Fig. 6a, the isospin effect of double ratios is washed out as shown in Fig.6b. 
The same picture was found experimentally in central collision of $^{112}$Sn + $^{112}$Sn and $^{124}$Sn + $^{124}$Sn  at 40 MeV/u and intepreted with the Expanding Emitting Source model \cite{Kund98}.
This independence on isospin  is consistent with the attainment of full chemical equilibrium for each disassembling source at the common temperature, which probably reflects that Albergo's temperature is insensitive to the isospin of the system. 

\subsection{Fragment Distribution and Related Critical Observables}
\subsubsection{Fragment Distribution}

In light of previous studies on the fragment distribution, we will use observables based on fragment to  signal the onset of phase transition, eg.  the effective power law
parameter $\tau$, the second moment $S_2$ of fragment distribution and the informtaion entropy. 
Before presenting these critical observables, the mass distribution of fragments are
shown at T = 4, 5, 6 and 7 MeV for $^{129}$Xe in Fig. 7. Clearly the disassembly mechanism evolves with the nuclear temperature. A few light clusters are emitted and the big residue reserves at T = 4 MeV which indicates typical evaporation mechanism. With the increasing  temperature, the shoulder of mass distribution of clusters occurs due to  the competition between the fragmentation and the evaporation. This shoulder disappears and the mass distribution becomes power law shape at T = 6 MeV, corresponding to  the multifragmentation region. When the temperature becomes much higher, the mass distribution becomes much steeper indicating that the disassembling process becomes more violent. Power law fits, Y(A) $\propto$ $A^{-\tau}$, for these mass distributions have been  introduced with the lines as shown in Fig.7.

\subsubsection{Effective Power Law Parameter}
It has already been observed that a minimum of power law parameter $\tau_{min}$ exists if the critical behavior takes place \cite{Fish67,Lit93}. Fig.8a displays $\tau$ parameter
as a function of temperature for Xe nuclei with the different isospin. The minimums of $\tau$ parameters in Fig.8a locate closely at 5.5 MeV for all the systems, which illustrates  
its minor dependence on the isospin. In  other words, there could be a universal
scaling for mass distribution regardless of the size of disassembling source when the critical phenomenon takes place. However, $\tau$ parameters show different values outside the critical region for nuclei with different isospin, eg., $\tau$ decreases with isospin when T $>$ 5.5 MeV (multifragmentation region).

\subsubsection{Campi's Second Moment}
In Fig.8b we give the temperature dependences of Campi's second moment of the fragment mass distribution \cite{Camp88}, which is defined as
\begin{equation}
S_2 = \frac{ \sum_{i \neq Amax} {A_i^2  n_i(A_i)}}{A} ,         
\end{equation}
where $n_i(A_i)$ is the number of clusters with $A_i$ nucleons and the 
sum excludes the largest cluster $A_{max}$, $A$ is the mass of the system.
At the percolation point $S_2$ diverges in an infinite system and is at maximum in
a finite system. Fig.8b gives the maximums of $S_2$ around 5.5 MeV for different isotopes, respectively. Again, the critical behavior occurs at the same temperature, independent of the isospin,  as $\tau$ reveals.

\subsubsection{Information Entropy}
Fig.8c plots the information entropy H as a function of temperature for Xe isotopes.  The information entropy was, for the first time, introduced by H. Shannon  in information theory  \cite{Denb85}. In high energy collisions, multiparticle production proceeds on the maximum stochasticity, $ie.$  they should obey the maximum entropy principal. This kind of stochasticity can be quantified via the information entropy. In the different physical condition  restrain, the information entropy can be defined with different stochastic variables. For examples, the information entropy for particle multiplicity can be defined as
\begin{equation}
 H = -\sum_{i} {p_i ln(p_i)}. 
\end{equation}
Here $p_i$ can be defined as an event  probability having $"i"$ produced particles,
the  sum is taken over all multiplicities of products from the disassembling system. $H$ reflects the capacity of the information or the extent of disorder.
Here we introduce this  entropy, for the first time,  into the search for liquid gas phase transition and critical phenomenon in the nuclear disassembly. Fig.8c shows that the entropy $H$ has peaks close to 5.5 MeV for all isotopes. These peaks indicate that the opening
of the phase space  in the critical point is the largest. In other words, the
system at the critical temperature has the largest fluctuation/stochasticity which leads to the largest disorder and particle production rate. Beyond the critical point, the information entropy $H$ increases with the isospin.

\subsubsection{Caloric Curve and Specific Heat}
Another direct quantities to illustrate  the phase transition are  caloric curve and specific heat. For an ideal infinite nuclear matter the first order liquid gas phase transition will reveal a temperature plateau in a wide range of excitation energy, which corresponds to an infinite specific heat in this constant temperature. But for a finite nuclear 
matter the plateau of caloric curve and the sharpness of specific heat will be largely smoothed.
But assumably, the specific heat might keep the imprint of the phase transition well and   probably reveal a salience with the temperature when phase transition and critical behavior holds. If exchanging the X and Y axis of Fig.1f, 
the  so-called caloric curves for Xe isotopes are obtained. For such finite nuclear systems,
caloric curves present that temperature increases generally with excitation energy
but a temperature plateau domain is always absent in this lattice gas model. On the other hand, at a fixed temperature, the excitation energy per nucleon decreases with isospin of dissociation isotopic sources, which is related to the bond numbers of unlike nucleons. Qualitatively, the higher the isospin of disassembling isotopic  source, the larger the $N_{np}^{T}$ and then the smaller the excitation energy per nucleon in terms of Eq. (2). Due to the caloric curves are not linear, the slopes of excitation energy to temperature for these curves, $ie.$ the specific
heat per nucleon at constant freeze-out volume (or density) $C_v/A$ for Xe isotopes is useful  to locate the critical point. Fig.3f shows the definite peaks of $C_v/A$ exist around 5 MeV for each disassembling sources, corresponding to the onset of the critical point behavior. Due to the influence of isospin on the  caloric curves, $C_v/A$ shows a similar anti-correlation with isospin especially below the critical temperature. 

\section{Summary}

In conclusion, the isospin effects of  particle emission and critical point behavior
 are explored for Xe isotopes at a fixed moderate freeze-out density in the frame of isospin dependent lattice gas model. Four sets of probe are testified to investigate such effect. First set of probe is the multiplicities of of LP, like $N_n$ and $N_p$, and the multiplicities of charged particles $N_{cp}$ and of intermediate mass fragments $N_{imf}$,
 and the average mass for the largest fragments  $A_{max}$ and the mean excitation energy per nucleon $E^*/A$; The second one is the ratios between two LP, namely the isotopic ratios, like $R(p/d)$ and $R(^3He/\alpha)$, or isobaric ratios, like  $R(t/^3He)$ and $R(^6He/^6Li)$, or  ratios of two light particls between which there exists one proton and mass number difference,  like $R(d/^3He)$ and $R(t/\alpha)$; The third one is the double ratios between two isotope-pairs, like $R(pd-He), R(pd-Li), R(dt-He), R(dt-Li)$, and $R(He-Li)$; The fourth one is the critical observables, like the effective power law parameter $\tau$, the information entropy $H$, 
 the Campi's second moment $S_2$, the caloric curves and the specific heat
 per nucleon $C_v/A$. The calculation illustrates that $N_n$, $N_p$ and ratios of light particles 
show the strong dependences on the isospin of the dissociation source, but it's not the case for the double ratios of light isotopes. The former reflects directly the chemical
composition of the source and the latter indicates the chemical equilibrium of the source, which is not self-contradict but reflects features of the sources on different sides.
Meanwhile, the critical temperature for a chain isotopes determined by the extreme values of $\tau$, $H$ and $S_2$ is also insensitive to the isospin of sources.  This conclusion is not contradict with the previous studies on the isospin dependence of critical temperature like in \cite{Gulm98}, where the span of isospin is from symmetrical nuclear matter to pure neutron matter. If we only look a small span of isospin for the experimentally measurable medium size isotopes, like $^{122-146}Xe$, the change of critical temperature is neglectable. This conclusion might indicate that it will be difficult to search for the isospin dependence of  critical temperature which signals the liquid gas phase transition  for medium size nuclei in the experimental point of view. In addition, values of 
power law parameter of cluster mass distribution, mean multiplicity of intermediate mass fragments ($IMF$), information entropy ($H$) and Campi's second moment ($S_2$)  also show minor dependence on the isospin of Xe isotopes at the critical point. In contrary, some isospin dependences of the values of $\tau$, $H$ and $S_2$ will reveal outside the critical region. Noting that the information entropy is, for the first time, introduced into the search for the liquid gas phase transition in this work.  Besides,  the slopes of  $N_n$, $N_p$, $N_{cp}$, $N_{imf}$, $A_{max}$ and $E^*/A$ to temperature and the variances of the distributions of these quantities are explored  as additional judgements for critical phenomenon. Similaring to the insensitivity of critical temperature to the isospin from the extreme values of $\tau$, $H$ and $S_2$, the critical temperature deduced from the above slopes and variances also shows independence on the isospin of the disassembling sources. It will be interesting to confront these conclusions with  the future experiments.

\vspace{0.5cm}
\centerline{ACKNOWLEDGEMENT}
\vspace{0.5cm}

The authors want to acknowledge Drs. Jicai Pan and S. Das Gupta for helps. This work was supported  by the National Nautural Science Foundation for Distinguished Young Scholar under Grant No. 19725521, National Natural Science Foundation of China under Grant No. 19705012,
Science and Technology Development Foundation of Shanghai under Grant No.
97QA14038, Presidential Foundation of Chinese Academy of Sciences,
and the Scientific Research Foundations for Returned Overseas Chinese
Scholar by the National Human Resource Administration, Education Administration
of China and Shanghai Government.



\figure{Fig.1: Average multiplicities for emitted neutrons (a), protons (b),
charged particles (c), intermediate mass fragments (d),
average values of the largest fragment masses (e) and of excitation energies per nucleon (f)
as functions of temperature and isospin of Xe.  }

\figure{Fig.2: The same as Fig.1 but as a function of excitation energy instead of temperature.}

\figure{Fig.3: Slopes of $N_n$ (a), $N_p$ (b), $N_{cp}$ (c), $N_{imf}$ (d)
and $A_{max}$ (e) to temperature. The symbol is the  same as in Fig.1} 

\figure{Fig.4: Variances for the distributions of $N_n$ (a), $N_p$ (b),
$N_{cp}$ (c), $N_{imf}$ (d), $A_{max}$ (e)  and $E^*/A$ (f) versus temperature.}

\figure{Fig.5: Isotopic ratios $R(p/d)$ (a) and  $R(d/t)$ 
 (b),  isobaric
ratios $R (t/^3He)$ (c) and $R(^6He/^6Li)$ (d), and ratios $R(d/^3He)$ (e)
and  $R(t/^4He)$ (f) as  functions of temperature and isospin of Xe. }

\figure{Fig.6: Iotopic ratios of $R(p/d)$, $R(d/t)$, $R(^3He/\alpha)$
and $R(^6Li/^7Li)$ (a) and double ratios $R(pd-He)$, $R(pd-Li)$,
$R(dt-He)$, $R(dt-Li)$ and $R(He-Li)$ (b) as a function of the isospin
 $(N-Z)/A$ around critical temperature $\sim$ 5 MeV.
 The symbols are illustrated in (a) and (b), respectively. }

\figure{Fig.7: Mass distribution of $^{129}$Xe at T = 4, 5, 6 and 7 MeV.
The lines are the power-law fit.}

\figure{Fig.8: Critical observables: $\tau$ parameter from the power law fit to mass distribution (a),  Campi's second moment  (b) and  information entropy (c)
as functions of temperature and isospin of disassembling sources.    }

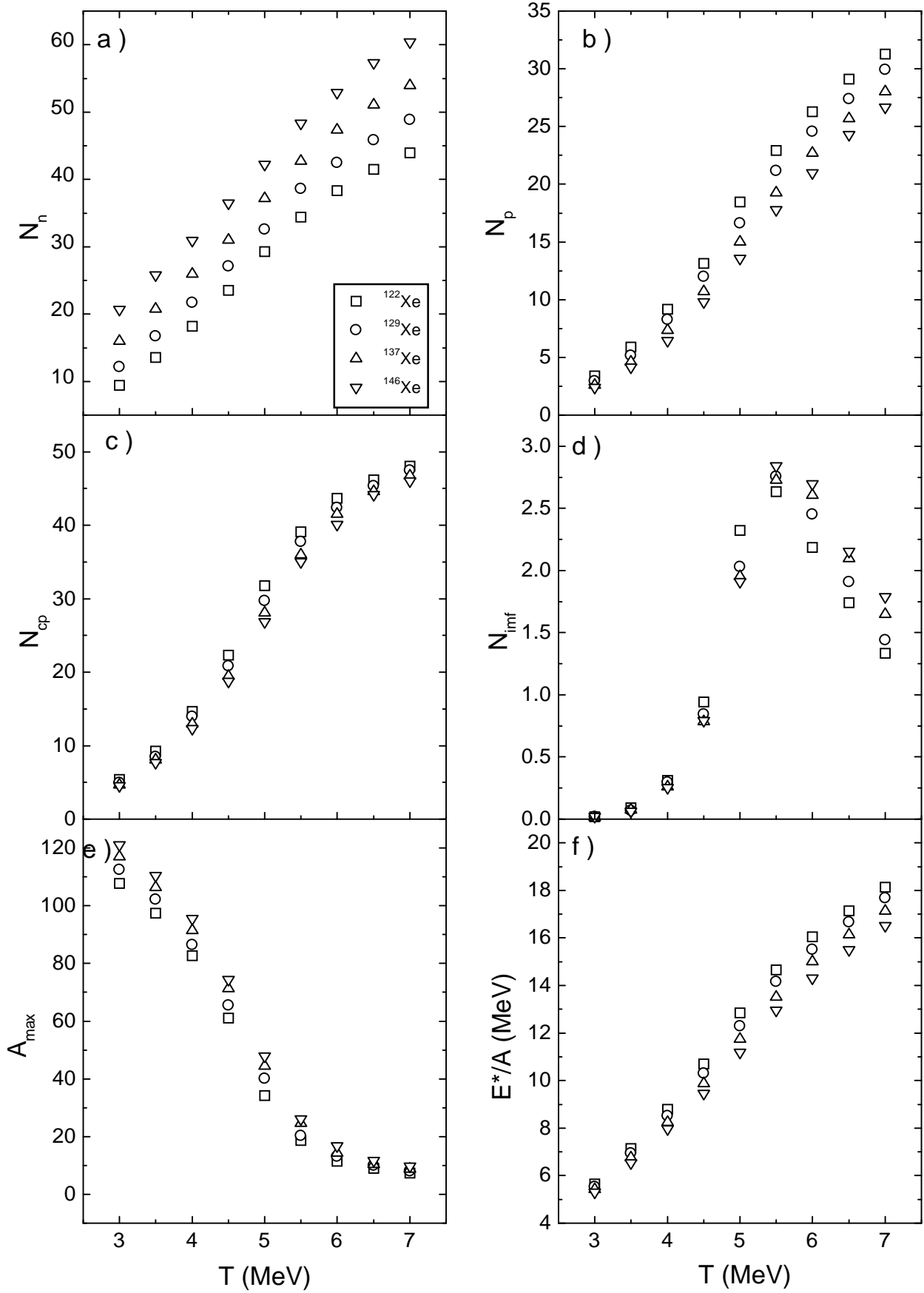
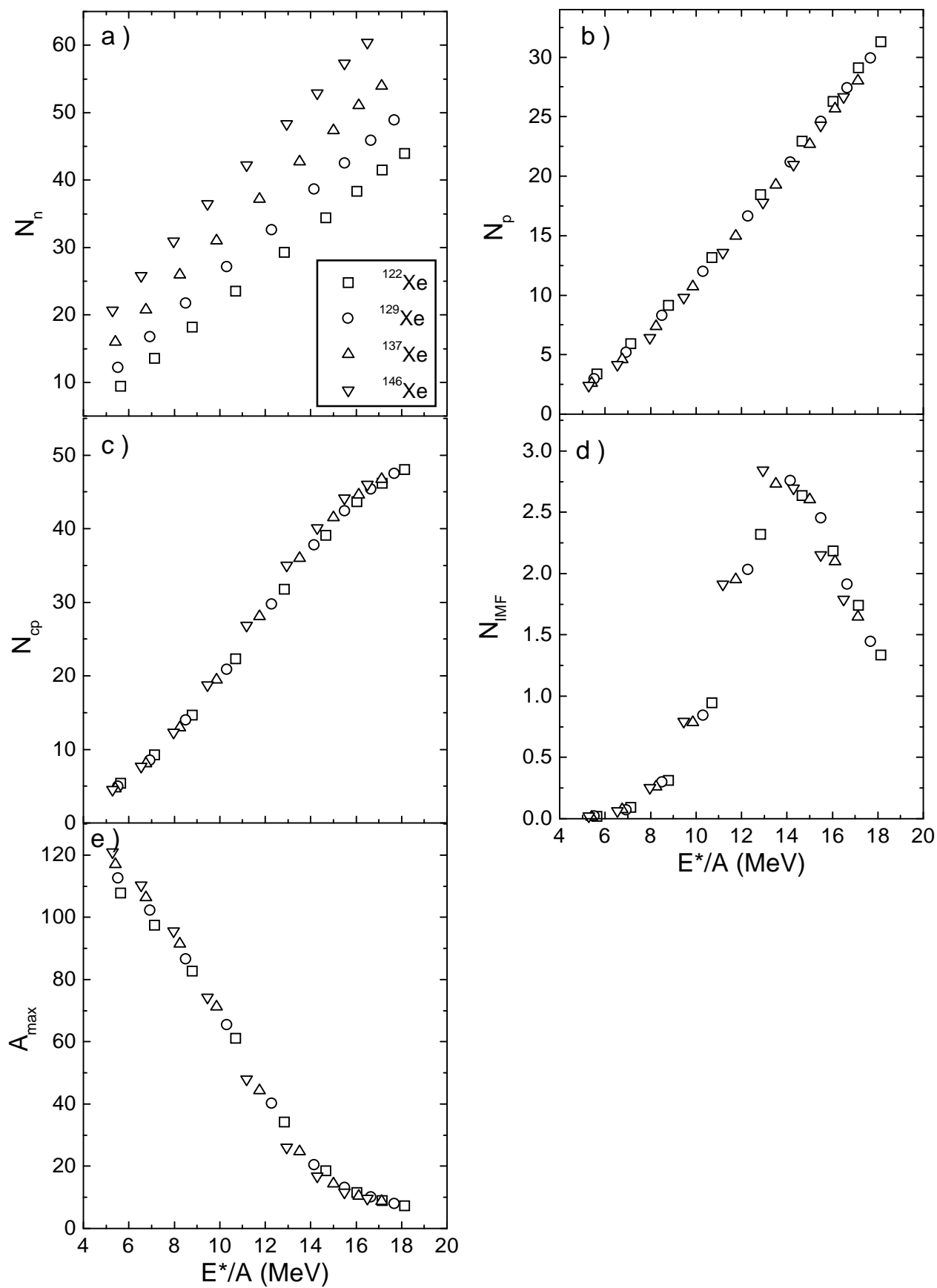
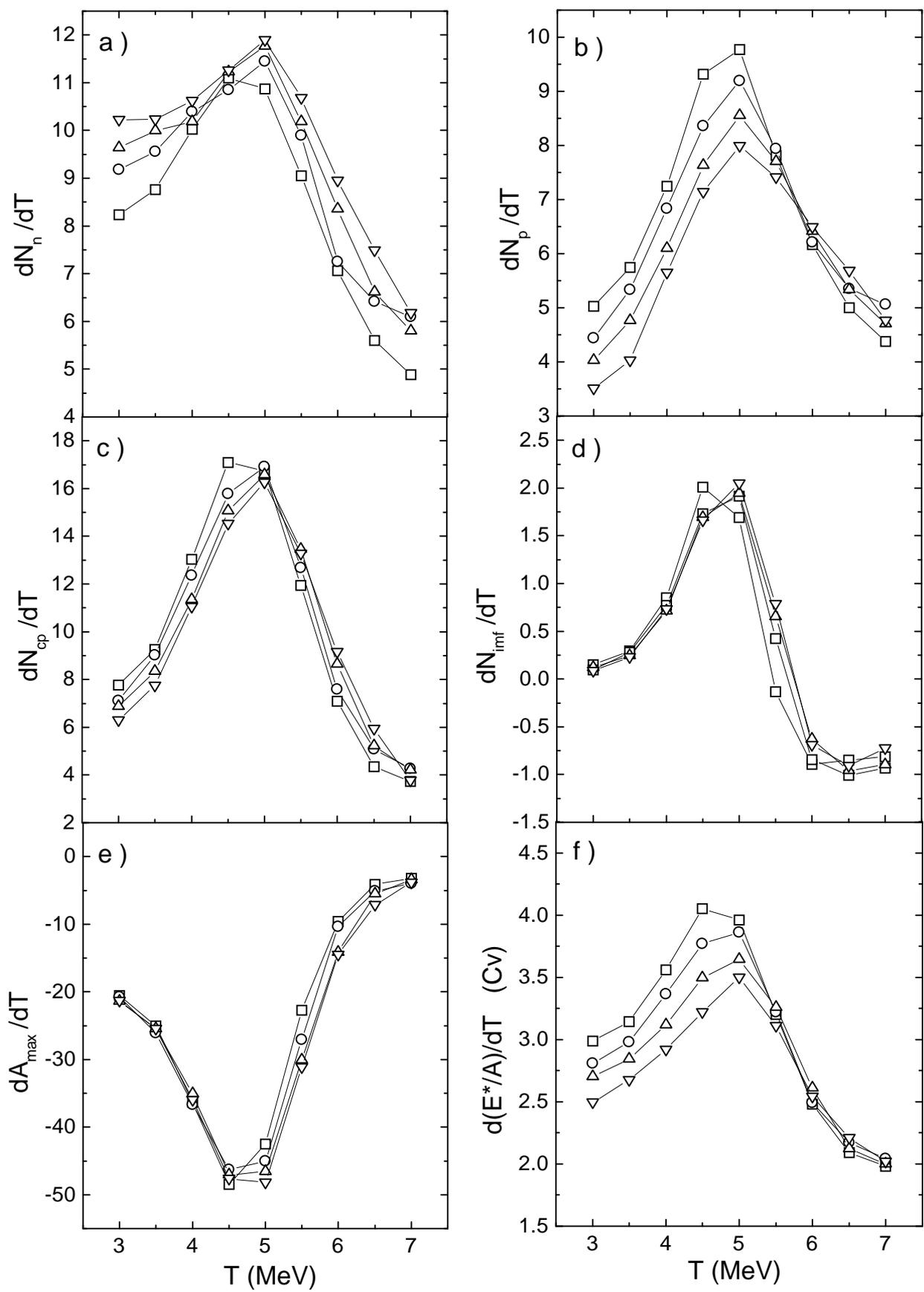
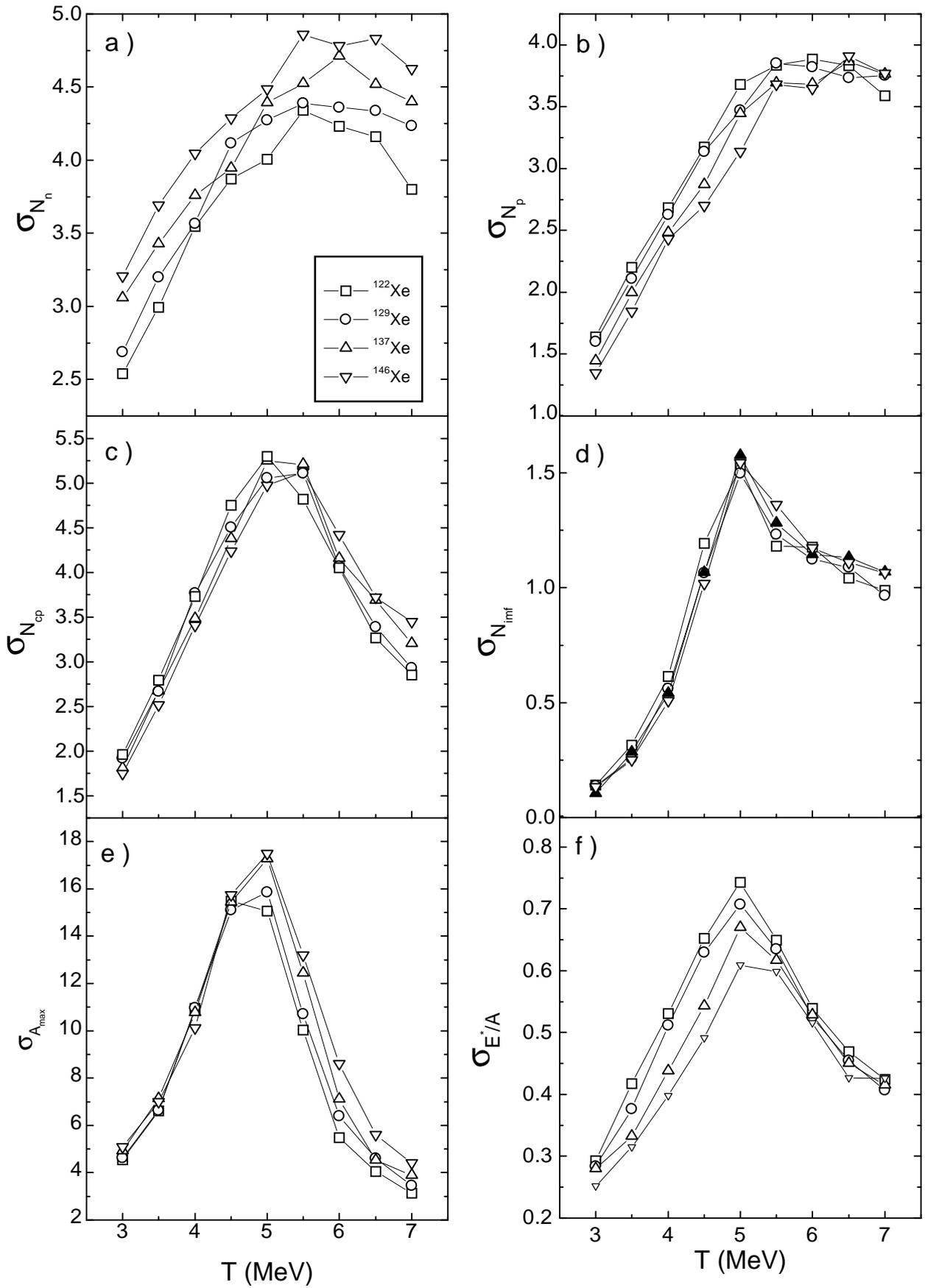
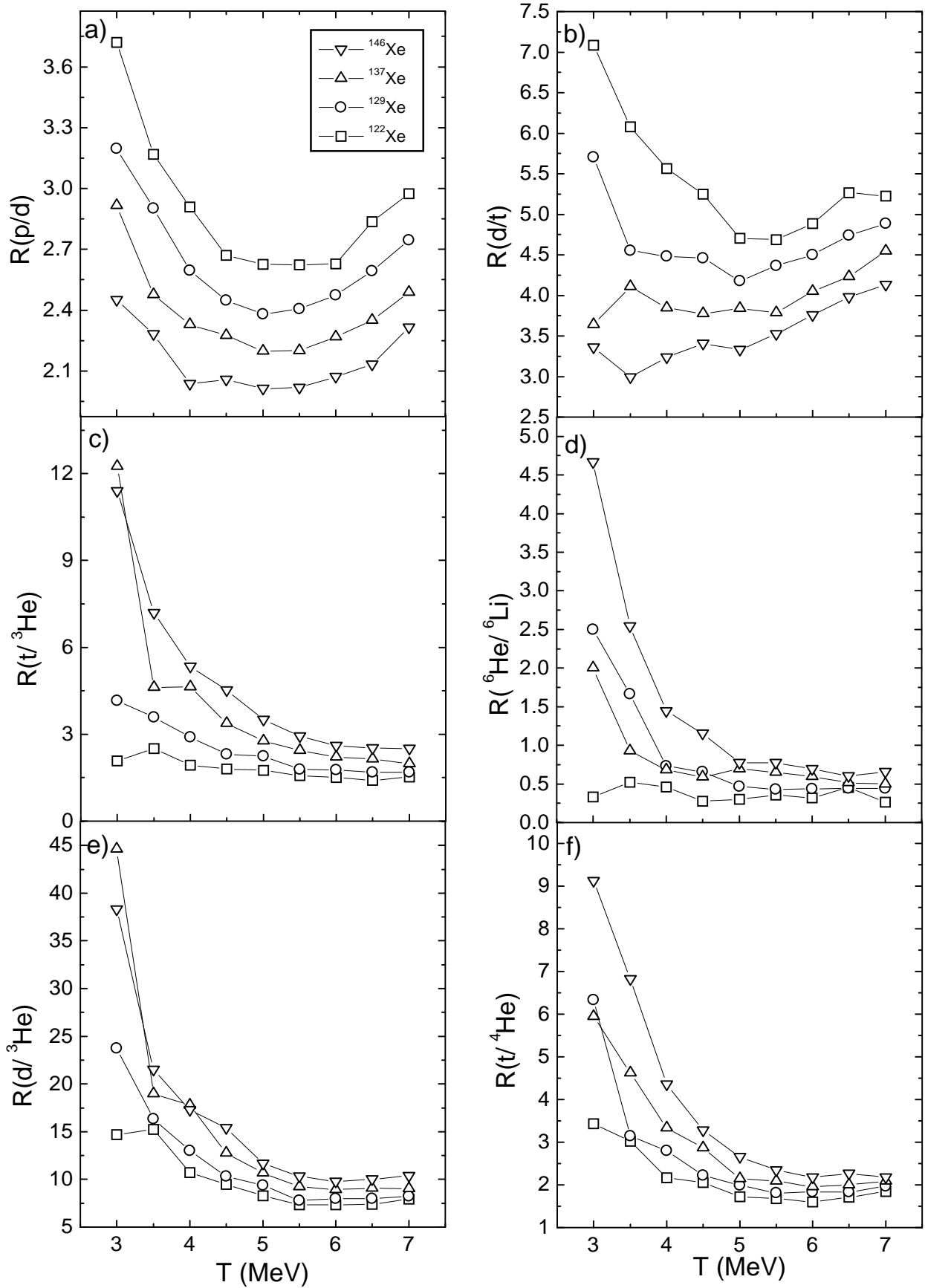
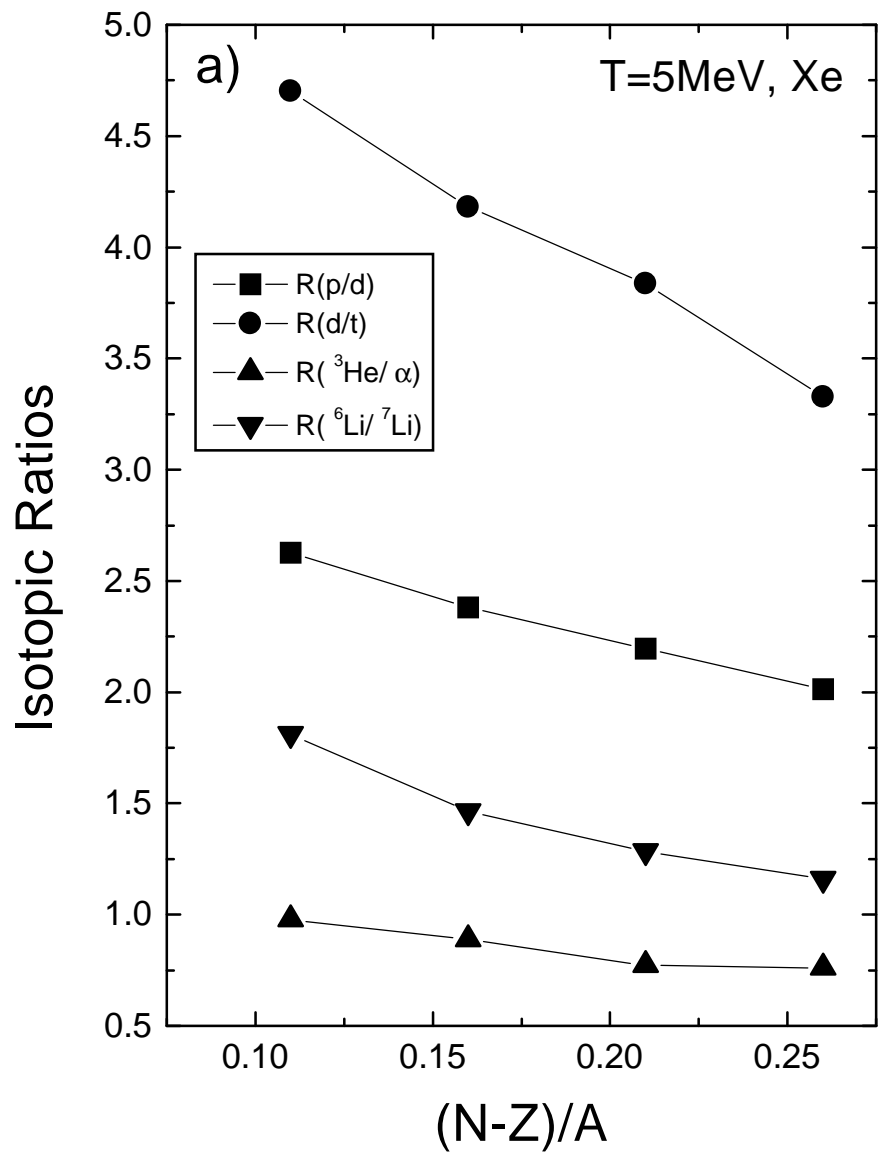
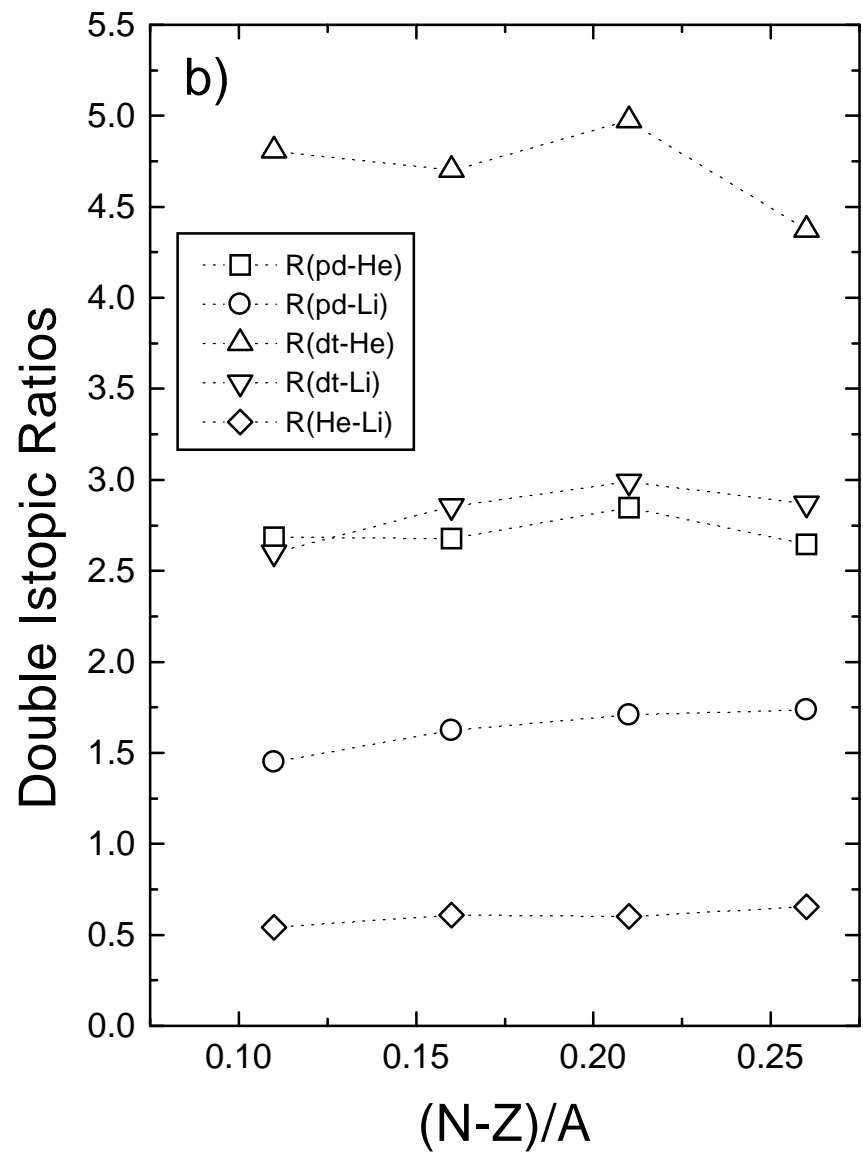
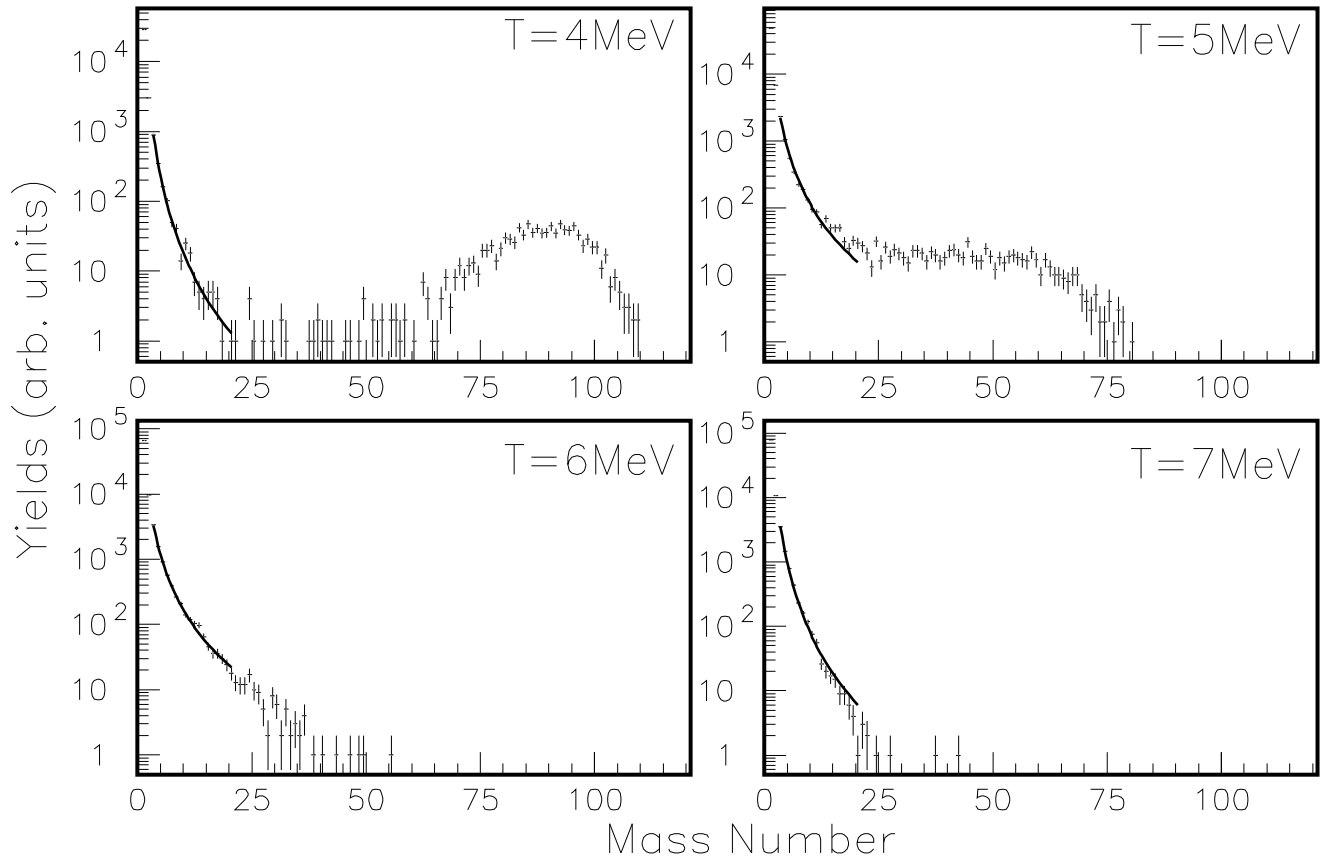
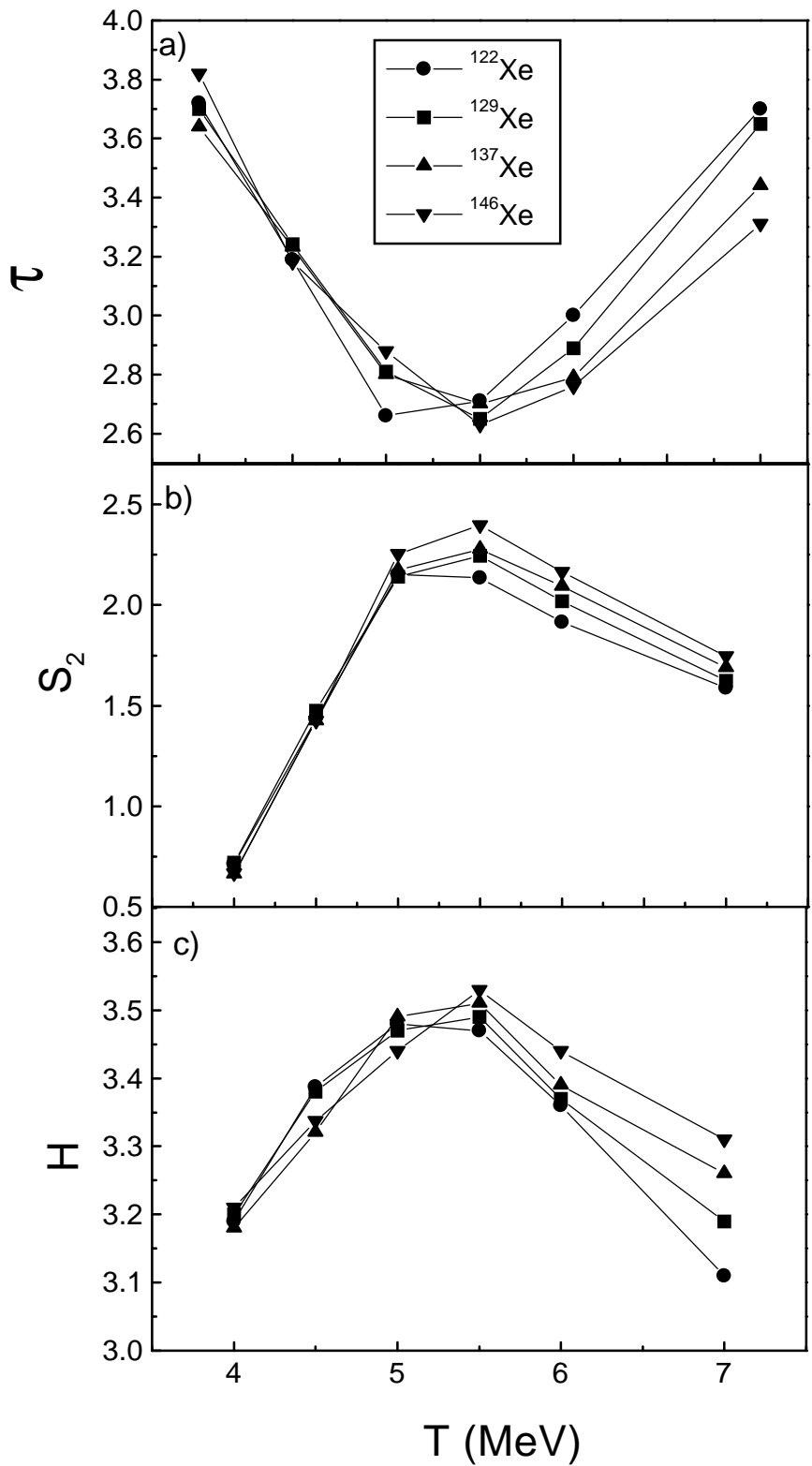
\end{document}